\documentclass[10pt,conference]{IEEEtran}
\usepackage{amsmath,amssymb,graphics,amsthm,psfrag,subfigure}
\usepackage{amsfonts,epsfig,graphicx,subfigure}
\usepackage{amsmath,amssymb,amsthm}
\usepackage{epic,eepic}
\usepackage{epsf}
\usepackage{fancyheadings}
\usepackage{graphics}
\usepackage{amsfonts}
\usepackage{amsmath}
\usepackage{amssymb}
\usepackage{epic}
\usepackage{eepic}
\usepackage{eepicemu}
\usepackage{psfrag}

\usepackage{arxivmacros}

\newcommand{\zlow}{\genericS{z}} 
 
\newcommand{\ylow}{\genericS{y}}

\newcommand{\rY}{\genericRV{Y}}

\newcommand{\mypara}[1]{\noindent {\bf{#1}}}

\makeatletter
\long\def\@makecaption#1#2{
        \vskip 0.8ex
        \setbox\@tempboxa\hbox{\small {\bf #1:} #2}
        \parindent 1.5em  %% How can we use the global value of this???
        \dimen0=\hsize
        \advance\dimen0 by -3em
        \ifdim \wd\@tempboxa >\dimen0
                \hbox to \hsize{
                        \parindent 0em
                        \hfil 
                        \parbox{\dimen0}{\def\baselinestretch{0.96}\small
                                {\bf #1.} #2
                                } 
                        \hfil}
        \else \hbox to \hsize{\hfil \box\@tempboxa \hfil}
        \fi
        }
\makeatother

\makeatletter
\long\def\barenote#1{
    \insert\footins{\footnotesize
    \interlinepenalty\interfootnotelinepenalty 
    \splittopskip\footnotesep
    \splitmaxdepth \dp\strutbox \floatingpenalty \@MM
    \hsize\columnwidth \@parboxrestore
    {\rule{\z@}{\footnotesep}\ignorespaces
      #1\strut}}}
\makeatother

\newcommand{\midbit}{\ensuremath{m}}

\newcommand{\myber}{\ensuremath{\operatorname{Ber}}}

\newcommand{\topbit}{\ensuremath{n}}
\newcommand{\lowbit}{\ensuremath{k}}
\newcommand{\mydefn}{\ensuremath{: \, =}}
\newcommand{\estim}[1]{\ensuremath{\widehat{#1}}}

\newcommand{\rval}{\ensuremath{t}}
\newcommand{\MomGen}[1]{\ensuremath{\mathbb{M}_{#1}}}
\newcommand{\lamstar}{\ensuremath{\lambda^*}}
\newcommand{\tmpq}{\ensuremath{u}}

\newcommand{\KeyFunc}{\ensuremath{F}}
\newcommand{\InterFunc}{\ensuremath{G}}

\newcommand{\ratediseff}[2]{\ensuremath{R_{\operatorname{up}}(#1; #2)}}
\newcommand{\ratediscom}[2]{\ensuremath{R_{\operatorname{com}}(#1; #2)}}

\newcommand{\WtEnumAsymp}{\ensuremath{\mathcal{A}}}
\newcommand{\rateldgm}{\ensuremath{R(\genMat)}}
\newcommand{\rateldpc}{\ensuremath{R(\parMat)}}

\newcommand{\rtotvar}{\ensuremath{\genericRV{U}}}
\newcommand{\rvaradd}{\ensuremath{\genericRV{Y}}}
\newcommand{\rvarplain}{\ensuremath{\genericRV{W}}}
\newcommand{\rcountvar}{\ensuremath{\genericRV{T}}}
\newcommand{\indber}[1]{\ensuremath{\inducedDmin{#1}}}
\newcommand{\IndBer}[1]{\indber{#1}}

\newcommand{\mybeginproof}{\noindent \emph{Proof: $\;$}}
\newcommand{\myendproof}{\hfill \qed}

\newcommand{\commonwidth}{0.36\textwidth}
\begin{document}

\title{Analysis of LDGM and compound codes for lossy compression and
binning}

\author{\authorblockN{Emin Martinian}
\authorblockA{Mitsubishi Electric Research Labs \\
Cambridge, MA  02139, USA \\
Email: martinian@merl.com}
\and
\authorblockN{Martin J. Wainwright}
\authorblockA{Dept. of Statistics and Dept. of EECS, \\
UC Berkeley,
Berkeley, CA  94720\\
Email: wainwrig@$\{$eecs,stat$\}$.berkeley.edu}}

\maketitle

%\comment{%% PREPRINT LABEL:
%\comment{
\vspace*{-2.8in}%
\vbox to 2.8in{\small\tt%
\begin{center}
  \begin{tabular}[t]{cl}
   Published in: &  Workshop on Information Theory and its Applications, \\
&  San Diego, CA.  February 2006
  \end{tabular} 
\end{center}
\vfil}
%}

\begin{abstract}
Recent work has suggested that low-density generator matrix (LDGM)
codes are likely to be effective for lossy source coding problems. We
derive rigorous upper bounds on the effective rate-distortion function
of LDGM codes for the binary symmetric source, showing that they
quickly approach the rate-distortion function as the degree increases.
We also compare and contrast the standard LDGM construction with a
compound LDPC/LDGM construction introduced in our previous work, which
provably saturates the rate-distortion bound with finite degrees.
Moreover, this compound construction can be used to generate nested
codes that are simultaneously good as source and channel codes, and
are hence well-suited to source/channel coding with side
information. The sparse and high-girth graphical structure of our
constructions render them well-suited to message-passing encoding.
\end{abstract}

\section{Introduction}

For channel coding problems, codes based on graphical constructions,
including turbo codes and low-density parity check (LDPC) codes, are
widely used and well understood~\cite{Richardson:it:2001}.  However,
many communication problems involve aspects of quantization, or
quantization in conjunction with channel coding.  Well-known examples
include lossy data compression, source coding with side information
(the Wyner-Ziv problem), and channel coding with side information (the
Gelfand-Pinsker problem).  For such communication problems involving
quantization, the use of sparse graphical codes and message-passing
algorithm is not yet as well understood.

A standard approach to lossy compression is via trellis-code
quantization (TCQ)~\cite{Marcellin90}, and various researchers have
exploited it for single-source and distributed
compression~\cite{Chou03,Yang05} as well as information embedding
problems~\cite{Chou01,Erez05}.  A limitation of TCQ-based approaches
is the fact that saturating rate-distortion bounds requires increasing
the trellis constraint length, which incurs exponential complexity
(even for message-passing algorithms).  It is thus of considerable
interest to explore alternative sparse graphical codes for lossy
compression and related problems.  A number of researchers have
suggested the use of LDGM codes for quantization
problems~\cite{martinian:2003:allerton,wainwright:2005:isit,Ciliberti05b,Murayama04}. Focusing
on binary erasure quantization (a special compression problem),
Martinian and Yedidia~\cite{martinian:2003:allerton} proved that LDGM
codes combined with modified message-passing can saturate the
fundamental bound.  A number of researchers have explored variants of
the sum-product algorithm~\cite{Murayama04} or survey propagation
algorithms~\cite{Ciliberti05a,wainwright:2005:isit} for quantizing
binary sources.  Suitably designed degree distributions yield
performance extremely close to the rate-distortion
bound~\cite{wainwright:2005:isit}.  Various researchers have used
techniques from statistical physics, including the cavity method and
replica methods, to provide non-rigorous analyses of LDGM performance
for source coding~\cite{Ciliberti05a,Ciliberti05b,Murayama04}.
However, thus far, it is only in the limit of zero-distortion that
this analysis has been made
rigorous~\cite{Creignou03,MezRicZec02,Cocco03,Dubois03}.

In this paper, we begin in Section~\ref{SecBoundsOne} by establishing
rigorous upper bounds on the effective rate-distortion function of
check-regular families of LDGM codes for all distortions $D \in [0,
\myhalf]$ under (maximum-likelihood) decoding.  Our analysis is based
on a combination of the second-moment method, a tool commonly used in
analysis of satisfiability problems~\cite{Creignou03,Dubois03}, with
standard large-deviation bounds.  Our bounds show that LDGM codes can
come very close to the rate-distortion lower bound.  Although the
residual gap vanishes rapidly as the check degrees are increased, it
remains non-zero for any finite degree.  In
Section~\ref{SecGenConstruc}, we discuss a LDPC/LDGM compound
construction, which we introduced in previous work~\cite{MarWai06a}.
Here we provide a refined analysis of the fact that this compound
construction can saturate the rate-distortion bound with finite
degrees.  We conclude in Section~\ref{SecDiscussion} with a discussion
of the extension of our constructions to source and channel coding
with side information~\cite{MarWai06b}, as well as the application of
practical message-passing algorithms~\cite{wainwright:2005:isit}.

\mypara{Notation:} Vectors/sequences are denoted in bold (\eg,
$\sSrc$), random variables in sans serif font (\eg, $\rvSrc$), and
random vectors/sequences in bold sans serif (\eg, $\rvsSrc$).
Similarly, matrixes are denoted using bold capital letters (\eg,
$\genMat$) and random matrixes with bold sans serif capitals (\eg,
$\rvGenMat$).  We use $I(\cdot;\cdot)$, $H(\cdot)$, and
$\rent{\cdot}{\cdot}$ to denote mutual information, entropy, and
relative entropy (Kullback-Leibler distance), respectively.  Finally,
we use $\cardinal{\{\cdot\}}$ to denote the cardinality of a set,
$\pNorm{\cdot}{p}$ to denote the $p$-norm of a vector, $\myber(t)$ to
denote a Bernoulli-$t$ distribution, and $\binent{t}$ to denote the
entropy of a $\myber(t)$ random variable.

\section{Bounds on standard LDGM constructions}
\label{SecBoundsOne}

In this section, we begin by defining the check-regular LDGM ensemble.
We then state and prove rigorous upper bounds on the effective
rate-distortion function of this ensemble under ML encoding.

\subsection{Check-regular ensemble and lossy compression}

\begin{figure}[h]
\begin{center}
\psfrag{#m#}{$\midbit$} \psfrag{#n#}{$\topbit$}
\psfrag{#topdeg#}{$\topdeg$} \psfrag{#G#}{$\genMat$}
\widgraph{.55\textwidth}{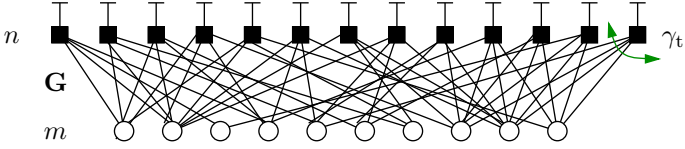}
\caption{Factor graph representation of an LDGM code with $\topbit$
checks (each associated with a source bit), and $\midbit$ information
bits.  The check-regular ensemble is formed by having each check
choose $\topdeg$ bit neighbors uniformly at random.}
\label{FigLDGM}
\end{center}
\end{figure}

A low-density generator matrix (LDGM) code of rate $\rate =
\frac{\midbit}{\numbit}$ consists of a collection of $\nbit$ checks
connected to a collection of $\midbit$ information bits; see
Figure~\ref{FigLDGM} for an illustration.
The ensemble of LDGM codes that we study in this paper are constructed
as follows: each check connects to $\topdeg$ information bits, chosen
uniformly and at random from the set of $\midbit$ information bits.
We use $\genMat \in \{0,1\}^{\midbit \times \topbit}$ to denote the
resulting generator matrix; by construction, each column of $\genMat$
has exactly $\topdeg$ ones, whereas each row (corresponding to a
variable node) has an (approximately) Poisson number of ones.  This
construction, while not particular good from the coding
perspective\footnote{In particular, for bounded check degree
$\topdeg$, the Poisson degree distribution means that there are
typically a constant fraction of isolated (degree zero) information
bits.}, has been studied in both the satisfiability and statistical
physics literatures~\cite{Creignou03,MezRicZec02,Cocco03,Dubois03},
where it is referred to as the ``$K$-XORSAT'' or ``$p$-spin'' model.
An advantage of this regular-Poisson degree ensemble is that the
resulting distribution of a random codeword is extremely easy to
characterize:
\begin{lems}
\label{LemInducedDmin}
Let $\rvGenMat \in \{0,1\}^{\midbit \times \topbit}$ be a random
generator matrix obtained by randomly placing $\topdeg$ ones per
column.  Then for any vector $\sIntm \in \{0,1\}^\midbit$ with a
fraction of $\weight$ ones, the distribution of the corresponding
codeword $\sIntm \, \rvGenMat$ is
Bernoulli($\inducedWeight{\weight;\topdeg}$) where
\begin{equation}
\label{eq:InducedWeight}
\inducedWeight{\weight;\topdeg} = \frac{1}{2} \cdot \left[1 - (1-2
  \weight)^{\topdeg}\right].
\end{equation}
\end{lems}

An LDGM code with generator matrix can be used to perform lossy data
compression as follows.  Given a source sequence $\ylow \in
\{0,1\}^\nbit$ drawn i.i.d. from a $\myber(\myhalf)$ source, we use it
to set the parities of the $\nbit$ checks at the top of
Figure~\ref{FigLDGM}.  We then seek an optimal encoding of the source
sequence by solving the optimization problem \mbox{$d(\estim{\ylow},
\ylow) \mydefn \min_{\zlow \in \{0,1\}^\midbit} d(\zlow' \genMat,
\ylow)$,} where $d(\cdot, \cdot)$ denotes the Hamming distortion.  For
a code of given rate $R$, we are interested in the expected minimum
distortion $\frac{1}{\topbit}\Exs[d(\estim{\rY}, \rY)]$ that can be
achieved, where the expectation is taken over the Bernoulli source.
For all distortions $\distor \in [0, \myhalf]$, the rate-distortion
function is well-known to take the form $R(\distor) = 1 -
\binent{\distor}$.

\subsection{Theoretical results}

\begin{figure}
\begin{center}
\widgraph{\commonwidth}{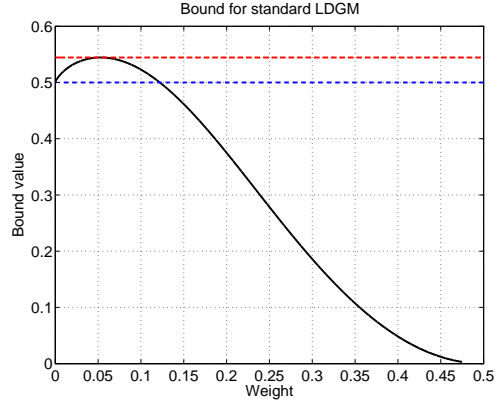}
\caption{Plot of the function $U(\weight; \distor, \topdeg)$ for
$\distor = 0.11$ and $\topdeg = 4$.  For $\weight = 0$, we have $U(0;
\distor; \topdeg) = 1 - \binent{\distor}$, so that the upper
bound~\eqref{EqnRateUpperBound} is always above the Shannon bound.
The value $\max_{\weight \in [0,1]} U(\weight; \distor, \topdeg)$
determines the excess rate required beyond the Shannon bound to
achieve distortion $\distor$. }
\label{FigRateGap}
\end{center}
\end{figure}
We begin by stating our main results on the rate-distortion
performance of LDGM codes.  For $\delta \in (0,1)$ and $\distor, \tmpq
\in [0, \myhalf]$, define $\lamstar(\delta, \distor, \tmpq) = \min\{0,
\log \rho^*(\delta, \distor, \tmpq) \}$, where $\rho^*$ is the unique
positive root\footnote{An explicit expression is $\rho^* = \frac{1}{2
A} \left[-B + \sqrt{B^2 -4 A C} \right]$.}  of the quadratic equation
$A x^2 + B x + C$ with coefficients
\begin{subequations}
\begin{eqnarray}
\label{EqnQuadRoots}
A & = & \delta \, (1-\delta) \, (1-\distor) \\
B & = & \tmpq (1-\delta)^2 + (1-\tmpq) \delta^2 - \distor
\left[\delta^2 + (1-\delta)^2 \right]. \\
C & = & - \distor \delta (1-\delta).
\end{eqnarray}
\end{subequations}
For $\delta = 0$, we set $\lamstar(0, \distor,\tmpq) = 0$.  Next
define the function $\KeyFunc[\distor, \delta]$ in a variational
manner as follows
\begin{multline}
\label{EqnDefnKeyFunc}
\max_{\tmpq \in [0,\distor]} \biggr \{ \binent{\tmpq} -
\binent{\distor} + \tmpq \log \left[(1-\delta) e^{\lamstar(\delta,
\distor,\tmpq)} + \delta\right] \\
+ (1-\tmpq) \log \left[ \delta e^{\lamstar(\delta, \distor, \tmpq)} +
(1-\delta) \right] - \distor \, \lamstar(\delta, \distor, \tmpq)
\biggr \}.
\end{multline}
With these definitions, we have:
\begin{theos}
\label{ThmRateDistBound}
The rate-distortion function of the $\topdeg$-regular ensemble is
upper bounded by
\begin{equation}
\label{EqnRateUpperBound}
\ratediseff{\distor}{\topdeg} \mydefn \max_{\weight \in [0, 1]} \left
\{ \frac{1 - \binent{\distor} + \KeyFunc \left[\distor;
\IndBer{\weight} \right]}{1 - \binent{\weight}} \right \}. \quad
\end{equation}
\end{theos}
To provide some intuition for the behavior of the function $U(\weight;
\distor, \topdeg) \mydefn \frac{1 - \binent{\distor} + \KeyFunc
\left[\distor; \IndBer{\weight} \right]}{1 - \binent{\weight}}$ that
determines the bound~\eqref{EqnRateUpperBound},
Figure~\ref{FigRateGap} provides a plot\footnote{Note that for even
$\topdeg$, the function $U$ is symmetric about $\myhalf$, so we only
plot one half of the function.} for the case $\distor = 0.11$ and
$\topdeg = 4$.  For $\weight = 0$, it can be seen that \mbox{$\KeyFunc
\left[\distor; \IndBer{0}\right] = 0$,} so that the \mbox{$U(0;
\distor, \topdeg) = 1 - \binent{\distor}$,} implying that the upper
bound is always larger than the Shannon lower bound.

\begin{figure}
\begin{center}
\widgraph{\commonwidth}{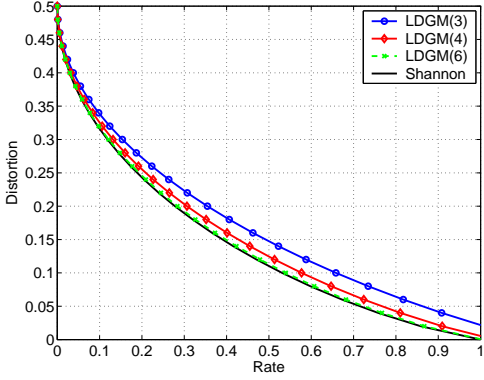} 
\caption{The Shannon rate-distortion function $\rate(\distor) =
1-\binent{\distor}$ provides a lower bound on any construction.  Plots
of the upper bound~\eqref{EqnRateUpperBound} for LDGM ensembles with
$\topdeg \in \{3,4,6\}.$ }
\label{FigPlots}
\end{center}
\end{figure}

By determining the maximum~\eqref{EqnRateUpperBound} for a range of
rates and degrees $\topdeg$, we can trace out parametric upper bounds
on the rate-distortion function.  Figure~\ref{FigPlots} provides plots
of the bound~\eqref{EqnRateUpperBound} on the rate-distortion function
for $\topdeg \in \{3, 4, 6\}$.  Also shown is the Shannon curve
$\rate(\distor) = 1 - \binent{\distor}$, which is a lower bound for
any construction.  Finally, an important special case of
Theorem~\ref{ThmRateDistBound} is the limit of zero distortion
($\distor = 0$), in which case the rate-distortion function
corresponds to the satisfiability threshold.  In this case, we recover
as a corollary the following result previously established by Creignou
et al.~\cite{Creignou03}:
\begin{cors}
\label{CorThreshold}
The random $\topdeg$-XORSAT satisfiability threshold is lower bounded
by $\alpha^*(\topdeg) \mydefn \frac{1}{\ratediseff{0}{\topdeg}}$,
where
\begin{eqnarray}
\label{EqnCorollary}
\ratediseff{0}{\topdeg} & = & \max_{\weight \in [0, \myhalf]} \frac{1
+ \log\left[1 - \IndBer{\weight} \right]}{1 - \binent{\weight}}
\end{eqnarray}
\end{cors}
This special case reveals that our upper bounds are not sharp, as the
bounds~\eqref{EqnCorollary} are known to be loose for the $\distor =
0$ case.  Indeed, several
researchers~\cite{MezRicZec02,Cocco03,Dubois03} have derived the exact
threshold values for the XORSAT problem.  However, the looseness in
the bound~\eqref{EqnCorollary} rapidly vanishes as $\topdeg$
increases.  As an illustration, for $\topdeg = 3$, we have
$\alpha^*(3) = 0.88949$ in contrast to the exact threshold $c^*(3) =
0.91794$, whereas for $\topdeg = 6$, we have $\alpha^*(6) = 0.99623$
in contrast to the exact threshold $c^*(6) = 0.99738$.

\subsection{Proof of Theorem~\ref{ThmRateDistBound}}

The remainder of this section is devoted to proving the previous
result.  Our proof exploits Shepp's second moment method, which is a
standard tool in satisfiability analysis:
\begin{lems}
\label{LemSecondMoment}
For any positive integer valued random variable $\rvz$, we have
$\Prob[\rvz > 0] \geq \frac{\E[\rvz]^2}{\E[\rvz^2]}$.
\end{lems}

Given an LDGM code $\fullCode$ of rate $\rate =
\frac{\midbit}{\topbit} > 0$, let $N = 2^{\topbit \rate}$ be the total
number of codewords.  For a given sequence $\sSrc \in
\{0,1\}^\topbit$, define for each codeword $i = 1, \ldots, N$ an
indicator variable $\indic{i}{\fullCode}{\sSrc}{D}$ for the event that
codeword $i$ is within Hamming distance $\distor \topbit$ of the
source sequence $\sSrc$.  Thus, the quantity
\begin{eqnarray}
\goodWords{\fullCode}{\sSrc}{\distor} & = & \sum_{i=1}^N
\indic{i}{\fullCode}{\sSrc}{\distor}
\end{eqnarray}
is the total number of codewords that are distortion
$\distor$-optimal.  In order to apply apply the second moment bound
(Lemma~\ref{LemSecondMoment}) to this random variable, we need to
compute the first and second moments.  Here we will be taking
expectations over both the source sequence $\sSrc \sim
\myber(\myhalf)$ and the choice of random code $\fullCode$ from the
$\topdeg$-regular ensemble.  In the following analysis, we will provide
conditions such that 
\begin{eqnarray*}
\log \Exs[\goodWords{\fullCode}{\sSrc}{\distor}]^2 -
\log \Exs[\goodWords{\fullCode}{\sSrc}{\distor}^2] & > &  -\log q(n),
\end{eqnarray*}
where $q(n)$ is a polynomial function of $n$.  It can be
shown~\cite{MarWai06a} using martingale arguments that such a
statement is sufficient to establish that the expected distortion is
less than $\distor$.  Consequently, we analyze normalized log
probabilities (i.e., $\frac{1}{\nbit} \log
\Exs[\goodWords{\fullCode}{\sSrc}{\distor}]$), and write $o(1)$ to
capture terms of the form $\frac{\log q(\nbit)}{\nbit}$.  The first
moment is straightforward to bound using standard results:
\begin{lems}
\label{LemFirstMoment}
The first moment is sandwiched as
\begin{subequations}
\begin{eqnarray}
\Exs[\goodWords{\fullCode}{\sSrc}{D}] & \geq & \frac{1}{\topbit+1}
2^{\topbit \left[ \rate - (1 - \binent{\distor})\right]}. \\
\Exs[\goodWords{\fullCode}{\sSrc}{D}] & \leq & (\topbit+1)\;
2^{\topbit \left[ \rate - (1 - \binent{\distor})\right]}.
\end{eqnarray}
\end{subequations}
\end{lems}
\noindent We also make use of the following alternative expression for
the second moment (see~\cite{MarWai06a} for a proof):
\begin{lems}
\label{LemSimpleSec}
The second moment can $\E[\rvGoodWords{D}^2]$ can be decomposed as
\begin{multline}
\label{EqnSimpleSec}
 E[\rvGoodWords{D}] + E[\rvGoodWords{D}] \; \sum_{j \neq 0} \Prob[
\rvIndic{j}{D}=1\mid\rvIndic{0}{D}=1].
\end{multline}
\end{lems}
\noindent Particularly important in our analysis is the following lemma,
which provides a large deviations upper bound on the conditional
probability in equation~\eqref{EqnSimpleSec}:
\begin{lems}
\label{LemKeyUpperBound}
Conditioned on the event that codeword $j$ has a fraction $\weight
\nbit$ ones, we have
\begin{eqnarray*}
\frac{1}{\topbit} \log \Prob\left [
\rvIndic{j}{D}=1\mid\rvIndic{0}{D}=1 \right] & \leq & \KeyFunc
\left[\distor; \IndBer{\weight} \right] + o(1), \quad
\end{eqnarray*}
where the function $\KeyFunc$ is defined in
equation~\eqref{EqnDefnKeyFunc}.
\end{lems}
\mybeginproof
We can reformulate the probability on the LHS as follows.  Let
$\rcountvar$ be a discrete variable with distribution
\begin{eqnarray*}
\Prob(\rcountvar = \rval) & = & \frac{{\nbit \choose
\rval}}{\sum_{s=0}^{\distor \nbit} {\nbit \choose s}} \qquad \mbox{for
$\rval = 0,1, \ldots, \distor \nbit$},
\end{eqnarray*}
representing the (random) number of $1$s in the source sequence
$\sSrc$.  Let $\rvaradd_i$ and $\rvarplain_j$ denote Bernoulli random
variables with parameters $1-\IndBer{\weight}$ and $\IndBer{\weight}$
respectively.  With this set-up, conditioned on codeword $j$ having a
fraction $\weight \nbit$ ones, the probability $\Prob [
\rvIndic{j}{D}=1\mid\rvIndic{0}{D}=1]$ is equivalent to the
probability that the random variable
\begin{eqnarray}
\label{EqnRandomFormulate}
\rtotvar & \mydefn & \sum_{i=1}^\rcountvar \rvaradd_j + \sum_{j=1}^{n -
\rcountvar} \rvarplain_j
\end{eqnarray}
is less than $\distor \nbit$.  To bound this probability, we use
Chernoff's bound in the form
\begin{eqnarray}
\label{EqnChernoff}
\frac{1}{\nbit} \log \Prob[\rtotvar\leq \distor \nbit] & \leq &
\inf_{\lambda < 0} \left( \frac{1}{\nbit} \log
\MomGen{\rtotvar}(\lambda) - \lambda \distor \right).
\end{eqnarray}
We begin by computing the moment generating function $\MomGen{\rtotvar}$.
Taking conditional expectations and using independence, we have
\begin{eqnarray*}
\MomGen{\rtotvar}(\lambda) & = & \sum_{\rval = 0}^{\distor \nbit}
 \Prob[\rcountvar = \rval] \; \left[\MomGen{\rvaradd}(\lambda)
 \right]^\rval \left[\MomGen{\rvarplain}(\lambda)\right]^{\nbit -
 \rval}
\end{eqnarray*}
Of interest to us is the exponential behavior of this expression in
$\nbit$.  Using the standard entropy approximation to the binomial
coefficient, we can write $\frac{1}{\nbit} \log
\MomGen{\rtotvar}(\lambda)$ as
\begin{multline*}
\frac{1}{\topbit} \log \biggr \{ \sum_{\rval=0}^{\distor \nbit} \exp
\big[ \nbit \big\{ \binent{\frac{\rval}{\nbit}} - \binent{\distor} +
\frac{\rval}{\nbit} \log \MomGen{\rvaradd}(\lambda) \\ +
\left(1-\frac{\rval}{\nbit}\right) \log \MomGen{\rvarplain}(\lambda)
\big \} \big] \biggr \} + o(1)
\end{multline*}
where the cumulant generating functions have the form
\begin{subequations}
\begin{eqnarray}
\log \MomGen{\rvaradd}(\lambda) & = & \log \left[(1-\delta) e^\lambda
+ \delta \right]. \\
\log \MomGen{\rvarplain}(\lambda) & = & \log \left[ (1-\delta) +
\delta e^\lambda \right].
\end{eqnarray}
\end{subequations}
Note that the exponential behavior of the Chernoff
bound~\eqref{EqnChernoff} is determined by $\max_{\tmpq \in [0,
\distor]} \InterFunc(\tmpq; \lambda)$ where
\begin{multline*}
\InterFunc(\tmpq; \lambda) \mydefn \binent{\tmpq} - \binent{\distor} +
\tmpq \log \MomGen{\rvaradd}(\lambda) \\
+  (1-\tmpq) \log
\MomGen{\rvarplain}(\lambda) - \lambda \distor.
\end{multline*}
Since cumulant generating functions are strictly convex, we are
guaranteed that $\InterFunc$ is strictly convex in $\lambda$;
similarly, it can be seen that $\InterFunc$ is strictly concave in
$\tmpq$.  Moreover, for any $\distor > 0$ and $\delta \in (0,1)$, we
have $\InterFunc(\tmpq; \lambda) \rightarrow +\infty$ as $\lambda
\rightarrow -\infty$. Thus, by standard min-max
results~\cite{Hiriart1}, we can interchange the order of minimization
(over $\lambda < 0$) and maximization (over $\tmpq \in [0,\distor]$).
Taking derivatives with respect to $\lambda$ to find the minimum, we
find that $\frac{\partial F}{\partial \lambda} = 0$ is equivalent to
\begin{equation*}
\tmpq \frac{(1-\delta) \exp(\lambda)}{(1-\delta) \exp(\lambda) +
\delta} + (1-\tmpq) \frac{\delta \exp(\lambda)}{(1-\delta) + \delta
\exp(\lambda)} - \distor \, = \, 0. \quad
\end{equation*}
This is a quadratic equation in $\exp(\lambda)$ with coefficients
specified in equation~\eqref{EqnQuadRoots}; the unique positive root
is $\rho^*$ as defined.  Finally, from the Chernoff
bound~\eqref{EqnChernoff}, we have
\begin{eqnarray*}
\frac{1}{\nbit} \log \Prob[ \rtotvar \leq \nbit \distor] & \leq &
\sup_{\tmpq \in [0, \distor]} \InterFunc(\tmpq; \lamstar(\tmpq;
\distor)).
\end{eqnarray*}
Recognizing that $\KeyFunc[\distor, \delta] = \sup_{\tmpq \in
[0,\distor]} \InterFunc(\tmpq; \lamstar(\tmpq; \distor))$ completes
the proof of the lemma.
\myendproof

We are now ready to complete the proof of the theorem.  First of all,
by combining Lemmas~\ref{LemFirstMoment} and ~\ref{LemKeyUpperBound},
we can upper bound $\frac{1}{\topbit} \log E[\rvGoodWords{D}] \;
\sum_{j \neq 0} \Prob[ \rvIndic{j}{D}=1\mid\rvIndic{0}{D}=1]$ by
\begin{equation*}
\rate \left(1-\binent{\distor} \right) + \max_{\weight \in [0, 1] }
\left \{ \rate \binent{\weight} + \KeyFunc[\distor, \IndBer{\weight}]
\right\}.
\end{equation*}
Combining with Lemma~\ref{LemSimpleSec}, we obtain that
$\frac{1}{\topbit} \log E[\rvGoodWords{D}^2]$ is upper bounded by
\begin{equation*}
\rate \left(1-\binent{\distor} \right) + \max_{\weight \in [0, 1] }
\left \{ \rate \binent{\weight} + \KeyFunc[\distor, \IndBer{\weight}]
\right\} + o(1).
\end{equation*}
Now plugging this bound into the second moment bound
(Lemma~\ref{LemSecondMoment}) and using Lemma~\ref{LemFirstMoment}, we
obtain that $\frac{1}{\topbit} \log \Prob[\rvGoodWords{D} > 0]$ is
lower bounded by
\begin{equation*}
\rate \left(1-\binent{\distor} \right) -\max_{\weight \in [0, 1] }
\left \{ \rate \binent{\weight} + \KeyFunc[\distor, \IndBer{\weight}]
\right\} + o(1).
\end{equation*}
The probability of finding a $\distor$-optimal word will \emph{not}
vanish exponentially fast as long as this quantity stays non-negative;
with some simple algebra, this condition is equivalent to the bound
\begin{eqnarray}
\rate & \geq & \max_{\weight \in [0, 1]} \frac{1 -
\binent{\distor} + \KeyFunc[\distor,
\IndBer{\weight}]}{1-\binent{\weight}}.
\end{eqnarray}
Therefore, the true rate distortion function must be smaller than the
RHS of this equation, thereby completing the proof of the theorem.
\myendproof

\subsection{Proof of Corollary~\ref{CorThreshold}}

To prove the corollary with $\distor = 0$, we note that
equation~\eqref{EqnRandomFormulate} now entails evaluating the
probability that $\sum_{j=1}^{\topbit} \rvarplain_j = 0$, where the
$\rvarplain_j$ are i.i.d. $\myber(\IndBer{\weight})$ variables.  By
Sanov's theorem, the error exponent (i.e., $\KeyFunc$) in this case is
simply $\rent{0}{\IndBer{\weight}} = \log(1-\IndBer{\weight})$.
Substituting this into equation~\eqref{EqnRateUpperBound} and using
the fact that $\binent{0} = 0$ yields the result.

\begin{figure}[h]
\begin{center}
\psfrag{#k#}{$\lowbit$}  \psfrag{#m#}{$\midbit$}
\psfrag{#topdeg#}{$\topdeg$} \psfrag{#vdeg#}{$\vdeg$}
\psfrag{#cdeg#}{$\cdeg$} \psfrag{#n#}{$n$} \psfrag{#G#}{$\genMat$}
\psfrag{#H1#}{$\parMat_1$} \psfrag{#H2#}{$\parMat_2$}
\widgraph{.51\textwidth}{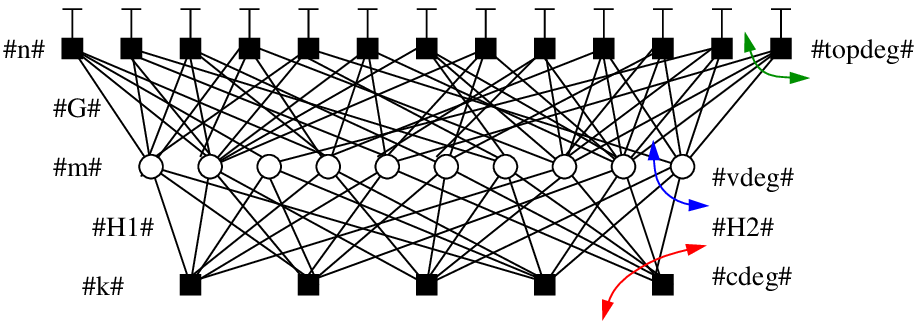}
\caption{Illustration of compound LDGM and LDPC code construction.
The top section consists of an $(\topbit, \midbit)$ LDGM code with
generator matrix $\genMat$ and constant check degrees $\topdeg = 4$;
its rate is $\rateldgm = \frac{\midbit}{\topbit}$.  The bottom section
consists of a $(\midbit, \lowbit)$ LDPC codes with degree $(\vdeg,
\cdeg) = (2,4)$, described by parity check matrix $\parMat$ and with
rate $\rateldpc = 1-\frac{\lowbit}{\midbit}$. }
\label{FigGenCompound}
\end{center}
\end{figure}

\section{Compound Constructions}
\label{SecGenConstruc}

In this section, we describe a compound construction, discussed in our
previous work~\cite{MarWai06a} in which an LDGM code is concatenated
with an LDPC code.  By contrast with the standard LDGM construction,
finite degrees suffice to saturate the rate-distortion bound.  The
compound code construction is illustrated in~\figref{FigGenCompound};
it is defined by a factor graph with three layers, and consists of an
LDGM code with generator matrix $\genMat$ and an LDPC code with parity
check matrix $\parMat$.  Note that a sequence $\ylow \in
\{0,1\}^\topbit$ is a codeword of this joint LDPC/LDGM construction if
and only if there exists an information sequence $\zlow \in
\{0,1\}^\midbit$ such that (a) $\zlow' \genMat = \ylow'$, and (b)
$\parMat \zlow = 0$ (where all operations are in modulo two
arithmetic).

\begin{figure}
\begin{center}
\widgraph{\commonwidth}{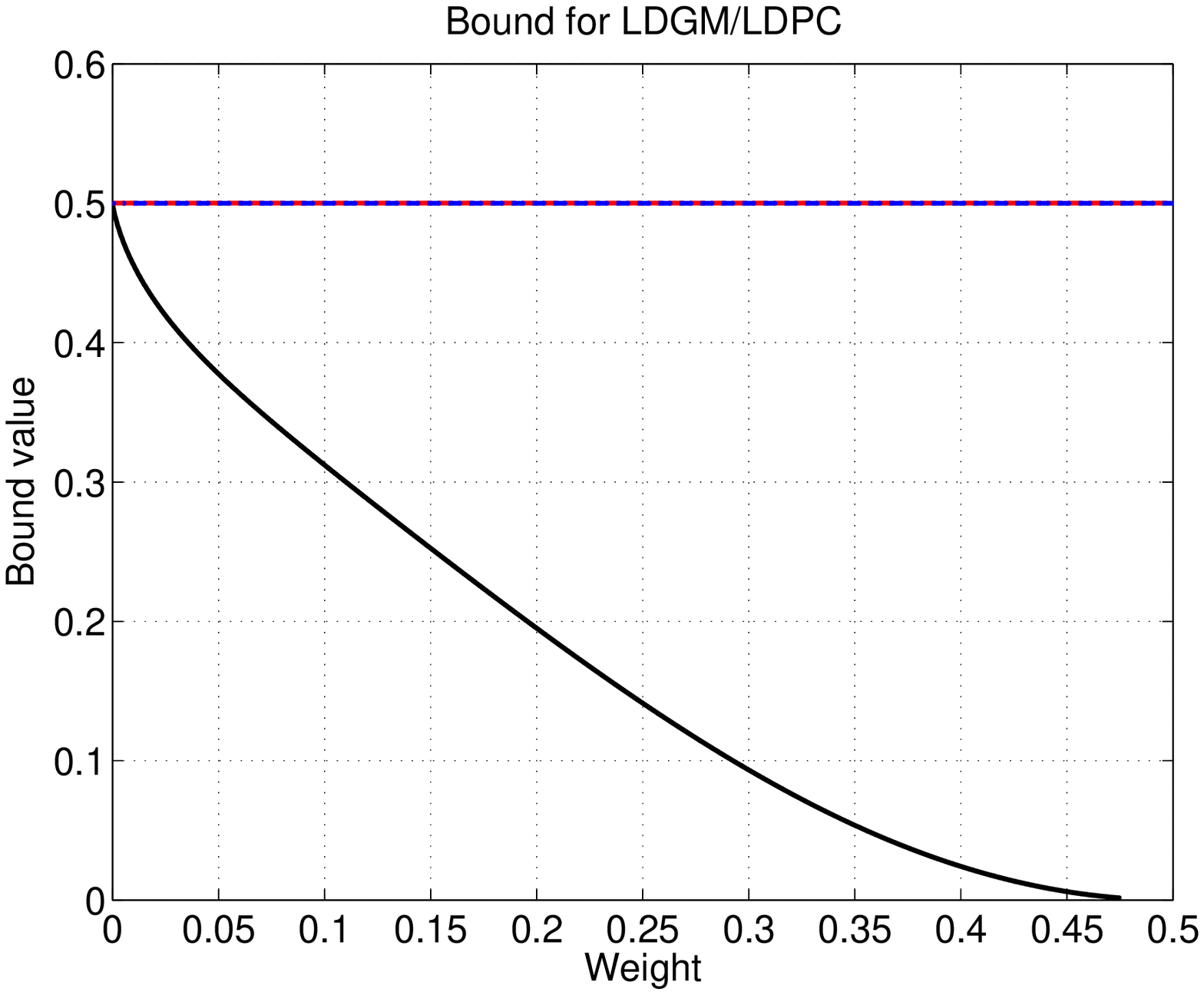}
\caption{Plot of the function $V(\weight; \distor, \topdeg)$ for
$\topdeg = 4$, a regular LDPC with degrees $(\vdeg, \cdeg) = (4,8)$,
rates $\rateldgm = 1$ and $\rateldpc = 0.5$, and distortion $\distor =
0.11$.  This function remains below $R = 0.5$ for all $\weight$, so
that the code saturates the Shannon lower bound.}
\label{FigNoOvershoot}
\end{center}
\end{figure}

The major deficit of LDGM codes---from the point of view of both
source and channel coding---is that they contain large numbers of
poorly separated codewords.  Herein lies the motivation for adding the
bottom LDPC precode: it serves to push apart the valid information bit
sequences $\zlow \in \{0,1\}^\midbit$, thereby spreading apart the
associated sequences $\zlow' \genMat$ that are codewords in the joint
LDGM/LDPC construction.  To formalize this intuition, a proof similar
to that of Theorem~\ref{ThmRateDistBound} establishes the following
\begin{theos}
\label{ThmGenBound}
The rate-distortion function of $\topdeg$-regular LDGM/LDPC compound
construction (with asymptotic LDPC weight enumerator
$\WtEnumAsymp(\weight)$) is upper bounded by
\mbox{$\ratediscom{\distor}{\topdeg} \mydefn \max_{\weight \in [0, 1]}
V(\weight; \distor, \topdeg)$,} where
\begin{equation}
\label{EqnGenRateUpperBound}
V(\weight; \distor, \topdeg) \mydefn \left \{ \frac{1 -
\binent{\distor} + \KeyFunc \left[\distor; \IndBer{\weight} \right]}{1
- \WtEnumAsymp(\weight) \big / \rateldpc} \right \}.
\end{equation}
\end{theos}
\noindent Note that this statement includes
Theorem~\ref{ThmRateDistBound} as a special case, in which $\rateldpc
= 1$ and $\WtEnumAsymp(\weight) = \binent{\weight}$.  Of interest to
us here is that these compound constructions (with $\rateldpc < 1$)
can saturate the rate-distortion bound with finite degrees.  The key
is that with suitable choice of LDPC degrees, we can ensure that
$\WtEnumAsymp(\weight)$ is negative in a region around zero, which
prevents the overshooting phenomenon illustrated in
Figure~\ref{FigRateGap}.  More specifically,
Figure~\ref{FigNoOvershoot} illustrates the analogous plot for a joint
LDGM/LDPC construction with $\topdeg = 4$, LDPC degrees $(\vdeg,
\cdeg) = (4,8)$, rates $\rateldgm = 1$ and $\rateldpc = 0.5$, and
distortion $\distor = 0.11$.
Notice how this curve remains below $R = 0.5$ for all $\weight \in [0,
0.5]$, demonstrating that the upper bound~\eqref{EqnGenRateUpperBound}
meets the Shannon lower bound.

\section{Discussion}
\label{SecDiscussion}

In concurrent work~\cite{MarWai06b}, we have shown that the joint
LDGM/LDPC construction in Figure~\ref{FigGenCompound} generates good
nested constructions (i.e., a good channel code can be partitioned
into good source codes, and vice versa), which can be shown to
saturate the Wyner-Ziv and Gelfand-Pinsker bounds.  We have also
shown~\cite{wainwright:2005:isit} that message-passing algorithms based
on survey propagation~\cite{Ciliberti05a}, when applied to LDGM codes
with suitable degree distributions, yield rate-distortion trade-offs
very close to the Shannon bound.  It remains to explore variants of
such message-passing algorithms for the compound construction, and
problems of coding with side information.

%%%%%%%%%%%%%%%%%%%%%%%%%%%%%%%%%%%%%%%%%%%%%%%%%%%%%%%%%%%%%%%%%%%

\section*{Acknowledgment}
{\small{ EM was supported by Mitsubishi Electric Research Labs and MJW
was supported by an Alfred P. Sloan Foundation Fellowship, an Okawa
Foundation Research Grant, and NSF Grant DMS-0528488.  The authors
thank Marc M\'{e}zard for helpful discussions.}}

\bibliographystyle{plain}
\bibliography{refs}

%%%%%%%%%%%%%%%%%%%%%%%%%%%%%%%%%%%%%%%%%%%%%%%%%%%%%%%%%%%%%%%%%%%%%%%%
\end{document}